\documentclass[12pt]{article}
\usepackage{epsf,amsfonts,latexsym}  
\newsymbol\ltimes 226E
\newsymbol\rtimes 226F
\epsfverbosetrue
\def\goth{\mathfrak}          
\def\double{\mathbb}

\def\cc{{\double C}}     
       
\def\rr{{\double R}}     
\def\zz{{\double Z}}

\def\aa{{\cal A}}
\def\ccc{{\cal C}}
 
\def\gg{{\goth g}}        
\def\hh{{\cal H}}

\def\aa{{\cal A}}
 
\def\hh{{\cal H}}

\def\lll{{\cal L}}

\def\t{{\rm tr}\,}

\def\ddd{{\,\hbox{$\partial\!\!\!/$}}}
\def\ddee{{\,\hbox{${\rm D}\!\!\!\!/\,$}}}

\def\de{\hbox{\rm d}} 

\def\pa{{\partial}}

\def\lb{\left[} 
\def\rb{\right]}

\def\op{\oplus}

\def\bb{\begin{eqnarray}}
\def\ee{\end{eqnarray}}
\def\eee{\nonumber\end{eqnarray}}
\def\pp{\pmatrix}
\def\qq{\quad}

\begin{document}

\hsize 17truecm
\vsize 24truecm
\font\twelve=cmbx10 at 13pt
\font\eightrm=cmr8
\baselineskip 18pt

\begin{titlepage}

\centerline{\twelve CENTRE DE PHYSIQUE TH\'EORIQUE}
\centerline{\twelve CNRS - Luminy, Case 907}
\centerline{\twelve 13288 Marseille Cedex 9}
\vskip 2truecm

\centerline{\twelve FORCES FROM NONCOMMUTATIVE GEOMETRY}

\bigskip

\begin{center} {\bf Thomas SCH\"UCKER}
\footnote{\, and Universit\'e de Provence,\qq
schucker@cpt.univ-mrs.fr } \\

\end{center}

\vskip 1truecm
\leftskip=1cm
\rightskip=1cm
\centerline{\bf Abstract} 

\medskip

 Einstein derived general relativity
from Riemannian geometry. Connes extends this derivation to
noncommutative geometry and obtains electro-magnetic, weak
and strong forces. These are pseudo forces, that accompany the
gravitational force just as in Minkowskian geometry the magnetic
force accompanies the electric force. The main physical input of
Connes' derivation is parity violation. His main output is the
Higgs boson which breaks the gauge symmetry spontaneously and
gives masses to gauge and Higgs bosons.

\medskip

 Einstein d\'eduit la gravitation \`a partir
de la g\'eom\'etrie Riemannienne. Connes \'etend cette
d\'erivation \`a la g\'eom\'etrie non commutative et obtient
les forces \'electro\-magn\'etique, faible et forte. Ce sont
des pseudo forces qui accompagnent la force gravitationnelle,
au m\^eme titre qu'en
g\'eom\'etrie Minkowskienne la force magn\'etique est une
pseudo force qui accompagne la force \'electrique. L'input
physique de la d\'erivation de Connes est la violation de la
parit\'e. Son r\'esultat principal est le scalaire de Higgs qui brise
la sym\'etrie de jauge  spontan\'ement et rend massifs les
bosons de jauge et de Higgs.

\vskip .5truecm PACS-92: 11.15 Gauge field theories\\ 
\indent MSC-91: 81T13 Yang-Mills and other gauge theories 
 
\vskip .5truecm

\noindent CPT-01/P.4245\\
\noindent hep-th/yymmxxx

 \end{titlepage}

Still today one of the major summits in physics is the
understanding of the spectrum of the hydrogen atom. The
phenomenological formula by Balmer and Rydberg was a
remarkable pre-summit on the way up. The true summit was
reached by deriving this formula from quantum mechanics. We
would like to compare the standard model of electro-magnetic,
weak and strong forces with the Balmer-Rydberg formula and
review the present status of Connes' derivation of this model from
noncommutative geometry. This geometry extends Riemannian
geometry and Connes' derivation is a natural extension of another
major summit in physics: Einstein's derivation of general
relativity from Riemannian geometry. Indeed, Connes' derivation
unifies gravity with the other three forces. 
 \begin{table}[h]
\begin{center}
\begin{tabular}{ll}
atoms&particles and forces\\[1ex]
Balmer-Rydberg formula \qq\qq&standard model\\[1ex]
quantum
mechanics&noncommutative geometry
\end{tabular}
\end{center}
\caption{An analogy}
\end{table}

Let us briefly recall four nested, analytic geometries and their
impact on our understanding of forces and time, see table 2. {\it
Euclidean geometry} is underlying Newton's mechanics as space of
positions. Forces are described by vectors living in the same space
and the Euclidean scalar product is needed to define work and
potential energy. Time is not part of geometry, it is absolute. This
point of view is abandoned in special relativity unifying space
and time into {\it Minkowskian geometry}. This new point of view
allows to derive the magnetic field from the electric field as a
pseudo force associated to a Lorentz boost. Although time has
become relative, one can still imagine a grid of synchronized
clocks, i.e. a universal time. The next generalization is Riemannian
geometry = curved spacetime. Here gravity can be viewed as the
pseudo force associated to a uniformly accelerated coordinate
transformation. At the same time universal time loses all meaning
and we must content ourselves with proper time. With today's
precision in time measurement, this complication of life becomes a
bare necessity, e.g. the global positioning system (GPS).

\begin{table}[h]
\begin{center}
\begin{tabular}{lll}
geometry&force & time\\[3ex]
Euclidean &$E=\int\vec F\cdot\de \vec x$&absolute\\[1ex]
Minkowskian&$\vec E,\epsilon _0\Rightarrow\vec B,\mu
_0=\,\frac{1}{\epsilon _0c^2}\, $&universal\\[1ex]
Riemannian&Coriolis $\leftrightarrow$ gravity&proper, $\tau$
\\[1ex] 
noncommutative\qq\qq&gravity $\Rightarrow$ YMH, $\lambda
={\textstyle\frac{1}{3}} g_2^2$\qq\qq&$\Delta \tau \sim 10^{-40}$ s
\end{tabular}
\end{center}
\caption{Four nested analytic geometries}
\end{table}

 Our last
generalization is to Connes' noncommutative geometry = curved
space(time) with uncertainty. It allows to understand some
Yang-Mills and some Higgs forces as pseudo-forces associated to
transformations, that extend the two coordinate transformations
above to the new geometry without points. Also, proper time comes
with an uncertainty. This uncertainty of some hundred Planck
times might be accessible to experiments through gravitational
wave detectors within the next ten years \cite{ac}.

\section{Slot machines and the standard model}

Today we have a very precise phenomenological description of
electro-magnetic, weak and strong forces. This description, the
standard model, works on a perturbative quantum level and, as
classical gravity, it derives from an action principle. Let us
introduce this action by analogy with the Balmer-Rydberg
formula.

\begin{figure}[h]
\label{slot1}
\epsfxsize=4cm
\hspace{5.5cm}
\epsfbox{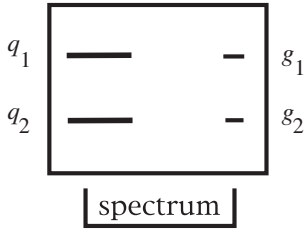}
\caption{A slot machine for atomic spectra}
\end{figure}

One of the new features of atomic physics was the appearance of
discrete frequencies and the measurement of atomic spectra
became a highly developed art. It was natural to label the discrete
frequences $\nu $ with natural numbers
$n$. To fit the spectrum of a given atom, say hydrogen, let us try
the ansatz
\bb\nu =g_1n_1^{q_1}+g_2n_2^{q_2}.\label{ansatz}\ee
We view this ansatz as a slot machine, you input two bills, that is
integers $q_1$, $q_2$ and two coins, that is two real numbers
$g_1$, $g_2$ and compare the output with the measured spectrum,
see figure 1. For the curious reader we should explain
why the integers
$n_j$ are considered more precious than the reals $g_j$. It is
because before Balmer and Rydberg there was a complicated
theory,  forgotten today, called exponent quantization. This theory
explained how --- assuming the existence of monopoles ---
exponents like the ones above are necessarily integers. Anyhow, if
you are rich enough you play and replay on the slot machine
until you win. The winner is the Balmer-Rydberg formula,
$n_1=n_2=-2$, $g_1=-g_2=\ 3.289\ 10^{15}$ Hz, the famous
Rydberg constant $R$. Then came quantum mechanics. It explained
why the spectrum of the hydrogen atom was discrete in the first
place, derived the exponents and the Rydberg constant,
\bb R=\,\frac{m_e}{4\pi \hbar^3}\,\frac{e_4}{(4\pi \epsilon
_0)^2}\, ,\ee
 from a noncommutativity,
$[x,p]=i\hbar 1$.

To cut short its long and complicated history we introduce the 
standard model as the winner of a particular slot machine. This
machine which has become popular under the names of Yang,
Mills and Higgs has
 four slots for four bills. Once you have decided which
bills you choose and entered them, a certain number of
small slots will open for coins. Their number depends
on the choice of bills. You make your choice of coins,
feed them in, and the machine starts working. It
produces as output a Lagrangian density. From this density
perturbative quantum field theory allows you to compute a
complete particle phenomenology: the particle spectrum with their
quantum numbers, cross sections, life times,
branching ratios, see figure 2. You compare the
phenomenology to experiment to find out whether your input
wins or loses.

\begin{figure}[h]
\label{slot2}
\epsfxsize=6cm
\hspace{4.7cm}
\epsfbox{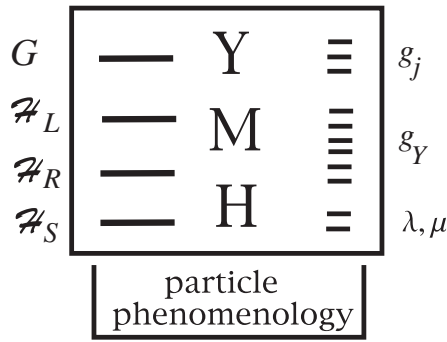}
\caption{The Yang-Mills-Higgs slot machine}
\end{figure}

\subsection{The input}

The first bill is a finite dimensional, real, compact Lie
group $G$. The gauge bosons $A$, spin 1, will live in its
adjoint representation whose Hilbert space is the
complexified of the Lie algebra $\gg$.

The remaining bills are three unitary
representations of $G$, $\rho_L,\ \rho_R$ and $ \rho_S$,
defined on the complex Hilbert spaces $\hh_L,\
\hh_R,\
\hh_S$. They classify the left- and right-handed
fermions, $\psi _L$ and $\psi _R$, spin ${\textstyle\frac{1}{2}}$,
and the scalars $\varphi $, spin 0. A massless
left-handed spinor has its spin parallel to its direction of
propagation, anti-parallel for the right-handed one. If the two
representations $\hh_L$ and $\hh_R$ are not identical, parity is
broken, because a space inversion, $\vec x\rightarrow -\vec x$
reverses the direction of propagation but leaves the spin
unchanged. The group
$G$ is chosen compact to ensure that the unitary representations
are finite dimensional, we want a finite number of `elementary
particles' according to the credo of particle physics that particles
are orthonormal basis vectors of the Hilbert spaces which carry
the representations. More generally, we might also admit
multi-valued representations, `spin representations' which would
open the debate on charge quantization. 

The coins are numbers, more
precisely coefficients of invariant polynomials. We 
need gauge couplings $g_j$ for each simple factor of $\gg$,
Then we need the Higgs potential $V(\varphi)$. It is
an invariant, fourth order, stable polynomial on
$\hh_S\owns\varphi$. Stable means bounded from
below. For $G=SU(2)\owns u$ and the Higgs scalar in the
fundamental representation,
$\varphi\in\hh_S=\cc^2$, $\rho_S(u)=u$, we have
\bb V(\varphi)=\lambda\,(\varphi^*\varphi)^2-
{\textstyle\frac{1}{2}}\mu^2\,\varphi^*\varphi.\ee
The coefficients of the Higgs potential are the Higgs
couplings, $\lambda$ must be positive for stability. We
say that the potential breaks $G$ spontaneously if no
minimum of the potential is invariant under $G$. In our
example, if $\mu$ is positive the minimum of 
$V(\varphi)$ is the 3-sphere $|\varphi|=v:=
{\textstyle\frac{1}{2}}\mu/\sqrt\lambda$. $v$ is
called vacuum expectation value and $SU(2)$ is said to
break down spontaneously.
 On the other
hand if $\mu$ is purely imaginary, then the
minimum of the potential is the origin, no
spontaneous symmetry breaking.
Finally, we need the Yukawa couplings $g_Y$. They
are the coefficients of the most general trilinear
invariant coupling between a scalar and two fermions, $\bar \psi
_L\varphi \psi _R$.

If the symmetry is broken
spontaneously, gauge and Higgs bosons acquire masses
related to gauge and Higgs couplings, fermions
acquire masses related to the Yukawa couplings.

The Lagrangian has five pieces, the Yang-Mills Lagrangian, the
Klein-Gordon Lagrangian, the Higgs potential, the Dirac
Lagrangian and the Yukawa terms:
\bb\lll\ =&{\textstyle\frac{1}{2g^2}} \t F_{\mu \nu }F^{\mu \nu
}+{\textstyle\frac{1}{2}} \overline{D_\mu \varphi }D^\mu
\varphi +V(\varphi )&\cr 
&+\bar{\psi } \ddee \psi 
+g_Y\bar{\psi } \varphi  \psi ,&\psi =\psi _L\op\psi _R.
\label{ymh}\ee
For $G=U(1)$ the Yang-Mills Lagrangian is nothing but Maxwell's
Lagrangian, the gauge boson $A$ is the photon and its coupling
constant $g$ is
$e\epsilon _0^{-1/2}$. The Dirac Lagrangian is the special
relativistic extension of Schr\"odinger's Lagrangian and $\psi$ is
the wave function of the electron and positron, coupled to the
electro-magnetic field $A$. Electro-magnetism preserves parity,
$\hh_L=\hh_R=\cc$, the representation being characterized by
the electric charge, $-1$ for both the left- and right handed
electron. The other three pieces are added by hand in order to give
masses to the gauge bosons and to the fermions. Without
spontaneous symmetry breaking such masses are forbidden by
gauge invariance and parity violation.

\subsection{The winner}\label{win}

Physicists have spent some thirty years and billions of
Swiss Francs playing on the slot machine by
Yang, Mills and Higgs. There is a winner, the standard
model of electro-weak and strong interactions. Its
bills are
\bb G&=&SU(2)\times U(1)\times
SU(3)/(\zz_2\times\zz_3),\label{group}\\ \cr 
\hh_L &=& \bigoplus_1^3\lb
(2,{\textstyle\frac{1}{6}},3)\op
(2,-{\textstyle\frac{1}{2}},1)
\rb  ,\label{hl}\\
 \hh_R& = &\bigoplus_1^3\lb 
(1,{\textstyle\frac{2}{3}},3)\oplus
(1,-{\textstyle\frac{1}{3}},3)\op (1,-1,1)
\rb,\label{hr} \\    
 \hh_S &= &(2,-{\textstyle\frac{1}{2}},1)\label{hs},
\ee   
where $(n_2, y, n_3)$
denotes the tensor product of an $n_2$ dimensional
representation of $SU(2)$, an $n_3$ dimensional
representation of $SU(3)$ and the one dimensional
representation of $U(1)$ with hypercharge $y$:  
$\rho(\exp (i\theta)) = \exp (iy\theta) $. For
historical reasons the hypercharge is an integer
multiple of ${\textstyle\frac{1}{6}}$.
This is irrelevant: only the product of the
hypercharge by its gauge coupling is measurable and we do not
need multi-valued representations which are characterized by
non-integer, rational hypercharges.
 In the direct sum, we recognize the three
generations of fermions, the quarks are $SU(3)$
colour triplets, the leptons colour singlets. The basis
of the fermion representation space is 
\bb \pp{u\cr d}_L,\ \pp{c\cr s}_L,\ \pp{t\cr b}_L,\ 
\pp{\nu_e\cr e}_L,\ \pp{\nu_\mu\cr\mu}_L,\ 
\pp{\nu_\tau\cr\tau}_L\eee
\bb\matrix{u_R,\cr d_R,}\qq \matrix{c_R,\cr s_R,}\qq
\matrix{t_R,\cr b_R,}\qq  e_R,\qq \mu_R,\qq 
\tau_R\eee
The
parentheses indicate isospin doublets. 

 We recognize the eight gluons in
$su(3)$. Attention, the $U(1)$ is not the one of electric
charge, it is called hypercharge, the electric charge
is a linear combination of hypercharge and weak
isospin, parameterized by the weak mixing angle
$\theta_w$ to be introduced below. This mixing is
necessary to give electric charges to the $W$ bosons.
The $W^+$ and $W^-$ are pure isospin states, while the
$Z^0$ and the photon are (orthogonal) mixtures of the
third isospin generator and hypercharge. 

 Because of the high
degree of reducibility in the bills, there are many
coins, among them 27 complex Yukawa couplings. Not
all of them have a physical meaning. The coins can be
converted into 18 physically significant, positive
numbers
\cite{data}, three gauge couplings, 
\bb g_1=0.3574\pm 0.0001,&g_2=0.6518\pm 0.0003,&
g_3=1.218\pm 0.01,\ee
two Higgs couplings, $\lambda$ and $\mu$, and 
13 positive parameters from the Yukawa couplings.
The Higgs couplings are related to the boson masses:
\bb m_W&=&{\textstyle\frac{1}{2}}g_2\,v
\,=\,80.419\pm 0.056\ {\rm GeV},\\
m_Z&=&{\textstyle\frac{1}{2}}\sqrt{g_1^2+g_2^2}\ v
=m_W/\cos\theta_w
\,=\,91.1882\,\pm\,.0022\ {\rm GeV},\\
m_H&=&2\sqrt 2\sqrt\lambda\,v\,>\,98\ {\rm GeV},\ee
with the vacuum expectation value
$v:={\textstyle\frac{1}{2}}\mu/\sqrt\lambda$ and the
weak mixing angle $\theta_w$ defined by
\bb\sin^2\theta_w:=g_2^{-2}/(g_2^{-2}+g_1^{-2})=
0.23117\,\pm\,0.00016.\ee
 For
the standard model, there is a one--to--one
correspondence between the physically relevant part
of the Yukawa couplings and the fermion masses and
mixings,
\bb m_e=0.510998902\pm 0.000000021\ {\rm MeV},&
m_u=3\pm 2\ {\rm MeV},&m_d=6\pm 3\ {\rm MeV},
\cr  
m_\mu=0.105658357\pm 0.000000005\ {\rm GeV},&
m_c=1.25\pm 0.1\ {\rm GeV},&
m_s=0.125\pm 0.05\ {\rm GeV},\cr  
m_\tau=1.77703 \pm 0.00003\ {\rm GeV},&
m_t=174.3\pm 5.1\ {\rm GeV},&
m_b=4.2\pm 0.2\ {\rm GeV}.\eee
For simplicity, we take massless neutrinos. Then  mixing only
occurs for quarks and is given by
a unitary matrix, the Cabibbo-Kobayashi-Maskawa
matrix 
\bb C_{KM}:=\pp{V_{ud}&V_{us}&V_{ub}\cr 
V_{cd}&V_{cs}&V_{cb}\cr  V_{td}&V_{ts}&V_{tb}}.\ee
For physical purposes it can be parameterized by
three angles $\theta_{12}$,  
$\theta_{23}$, $\theta_{13}$ and
one $CP$ violating phase $\delta$:
\bb C_{KM}=\pp{
c_{12}c_{13}&s_{12}c_{13}&s_{13}e^{-i\delta}\cr 
-s_{12}c_{23}-c_{12}s_{23}s_{13}e^{i\delta}&
c_{12}c_{23}-s_{12}s_{23}s_{13}e^{i\delta}&
s_{23}c_{13}\cr 
s_{12}s_{23}-c_{12}c_{23}s_{13}e^{i\delta}&
-c_{12}s_{23}-s_{12}c_{23}s_{13}e^{i\delta}&
c_{23}c_{13}},\ee
with $c_{kl}:=\cos \theta_{kl}$, 
$s_{kl}:=\sin \theta_{kl}$.
The
absolute values of the matrix elements are:
\bb \pp{
0.9750\pm 0.0008&0.223\pm 0.004&0.004\pm
0.002\cr 
0.222\pm 0.003&0.9742\pm 0.0008&0.040\pm 0.003\cr 
0.009\pm 0.005&0.039\pm 0.004&0.9992\pm 0.0003}.
\eee
The physical meaning of the quark mixings is the
following: when a sufficiently energetic $W^+$ decays
into a $u$ quark, this
$u$ quark is produced together with a
$\bar d$ quark with probability $|V_{ud}|^2$, together 
with a
$\bar s$ quark with probability $|V_{us}|^2$, together
with a
$\bar b$ quark with probability $|V_{ub}|^2$. The
fermion masses and mixings together are an entity,
the fermionic mass matrix or the matrix of Yukawa
couplings multiplied by the vacuum expectation
value. 

Let us note
six intriguing properties of the standard model.
\begin{itemize}\item
The gluons couple in the same way to left- and
right-handed fermions, the gluon coupling is
vectorial, strong interaction do not break parity.
\item
The fermionic mass matrix commutes with $SU(3)$, the three
colours of a given quark have the same mass.
\item
 The scalar is a colour singlet, the
$SU(3)$ part of $G$ does not suffer spontaneous break
down, the gluons remain massless.
\item
The $SU(2)$ couples only to left-handed fermions, its
coupling is chiral, weak interaction break parity
maximally.
\item
The scalar is an isospin doublet, the $SU(2)$ part
suffers spontaneous break down, the $W^\pm$ and the
$Z^0$ are massive.
\item
The remaining colourless and neutral gauge boson, the photon, is
massless and couples vectorially. This is certainly the most ad-hoc
feature of the standard model. Indeed the photon is a linear
combination of isospin which couples only to left-handed
fermions and of a $U(1)$ generator, that may couple to both
chiralities. Therefore only the
careful fine tuning of the hypercharges in the three input
representations (\ref{hl}-\ref{hs}) can save parity conservation
of electro-magnetism.
\end{itemize}
Nevertheless the phenomenological success of the standard model
is phenomenal: with only a handful of parameters it reproduces
correctly some millions of experimental numbers. And so far the
standard model is uncontradicted.

Let us come back to our analogy between the Balmer-Rydberg
formula and the standard model. One might object that the ansatz
for the spectrum, equation (\ref{ansatz}), is completely ad hoc,
while the class of all (anomaly free) Yang-Mills-Higgs models is
distinguished by perturbative renormalizability. This is true, but
this property was proved \cite{renorm} only years after
the electro-weak part of the standard model was published
\cite{gsw}.

By placing the hydrogen atom in an electric or magnetic field we
know experimentally that every frequency `state'
$n$, $n=1,2,3,...$ comes with $n$ irreducible unitary
representations
$\ell$,
$\ell=0,1,2,...n-1$ of dimensions $2\ell+1$. An orthonormal basis of
each representation
$\ell$ is labelled by another integer $m$, $m=-\ell,-\ell+1,...\ell$.
This experimental fact has motivated the credo that particles are
orthonormal basis vectors of unitary representations of compact
groups. This credo is also behind the standard model. While $SO(3)$
has a clear geometrical interpretation, we are still looking for
such an interpretation of $SU(2)\times U(1)\times
SU(3)/[\zz_2\times\zz_3].$

We close this section with Iliopoulos' joke from 1976
\cite{joke}. Meanwhile his joke has become hard, experimental
reality:

\medskip
\noindent{\bf Do-it-yourself kit for gauge models:}
\begin{enumerate}
\item[\bf 1)]
Choose a gauge group $G$.
\item[\bf 2)]
Choose the fields of the ``elementary particles'' you want to
introduce, and their representations. Do not forget to include
enough fields to allow for the Higgs mechanism.
\item[\bf 3)]
Write the most general renormalizable Lagrangian invariant
under $G$. At this stage gauge invariance is still exact and all
vector bosons are massless.
\item[\bf 4)]
Choose the parameters of the Higgs scalars so that spontaneous
symmetry breaking occurs. In practice, this often means to choose
a negative value for the parameter $\mu ^2$.
\item[\bf 5)]
Translate the scalars and rewrite the Lagrangian in terms of the
translated fields. Choose a suitable gauge and quantize the theory.
\item[\bf 6)]
Look at the properties of the resulting model. If it resembles
physics, even remotely, publish it.
\item[\bf 7)]
GO TO \bf 1.
\end{enumerate}

\section{Connes' noncommutative geometry}

Connes equips Riemannian spaces with an uncertainty principle.
As in quantum mechanics, this uncertainty principle derives from
noncommutativity.

Consider the classical harmonic oscillator. Its phase space is
$\rr^2$ with points labeled by position $x$ and momentum $p$. A
classical observable is a differentiable function on phase
space, for example the total energy $p^2/(2m)\,+\,kx^2$.
Observables can be added and multiplied, they form the algebra
$\ccc^\infty(\rr^2)$ which is associative and commutative. To pass
to quantum mechanics, this algebra is rendered noncommutative
by means of the following noncommutation relation for the
generators
$x$ and $p$,
\bb [x,p]=i\hbar 1.\ee
Let us call $\aa$ the resulting algebra `of quantum observables'. It
is still associative, has an involution $\cdot^*$, the adjoint, and a
unit, 1. Of course there is no space anymore of which
$\aa$ is the algebra of functions. Nevertheless we talk about such a
`quantum phase space' as a space that has no points or a space with
an uncertainty relation. Indeed the noncommutation relation
implies Heisenberg's uncertainty relation
\bb \Delta x\Delta p \geq \hbar /2\ee
and tells us that points in phase space lose all meaning, we can
only resolve cells in phase space of volume $\hbar/2$.

To define the uncertainty $\Delta a$ for an observable $a\in\aa$
we need a faithful representation of the algebra on a Hilbert
space,  i.e. an injective homomorphism
$\rho :\aa\rightarrow {\rm End}(\hh)$. 
For the harmonic oscillator this Hilbert space is
$\hh=\lll^2(\rr)$. Its elements are the wave functions $\psi (x)$,
 square integrable functions on configuration space. Finally
the dynamics is defined by a self adjoint observable $H=H^*\in\aa$
via Schr\"odinger's equation
\bb \left( i\hbar \,\frac{\pa}{\pa t}\, -\,\rho (H)\right) \psi
(t,x)=0.\ee
Usually the representation is not written explicitly. Since it is
faithful no confusion should arise from this abuse. Here time is
considered an external parameter, in particular time is not
considered an observable. This is different in the special
relativistic setting where Schr\"odinger's equation is replaced by
Dirac's equation,
\bb\ddd\psi =0.\ee
Now the wave function $\psi $ is the four component spinor
consisting of left- and right-handed, particle and antiparticle
wave functions.  The Dirac operator is not in
$\aa$ anymore, but
$\ddd\in{\rm End}(\hh)$. It is formally self adjoint,
$\ddd^*=\ddd.$

Connes' geometries are described by these three purely algebraic
items, $(\aa,\hh,\ddd)$, with $\aa$ a real, associative, possibly
noncommutative involution algebra with unit, faithfully
represented on a complex Hilbert space $\hh$ and $\ddd$ is a
self adjoint operator on $\hh$.

Connes'  geometry \cite{book}
 does to spacetime what quantum mechanics does
to phase space. So the first question we have to ask is: can we 
reconstruct Riemannian geometry from the algebraic data of
the so called {\bf spectral triple}
$(\aa,\hh,\ddd)$. The answer is affirmative precisely in the case
where the algebra $\aa$ is commutative. Indeed Connes'
reconstruction theorem of 1996 \cite{grav} establishes a
one-to-one correspondence between commutative spectral triples
and Riemannian spin manifolds.

 Let us try to get a feeling of the local
information contained in this theorem. Besides describing the
dynamics of the spinor field $\psi $ the Dirac operator $\ddd$
encodes the Riemannian metric, which
is the gravitational field, and the dimension of spacetime can be
read from its spectrum. The square of the Dirac operator is the
wave operator which in 1+2 dimensions governs the dynamics of a
drum. Remember the question `Can you hear the shape of a
drum?' that relates physics and mathematics in a beautiful way.
This question concerns a global property of spacetime, the
boundary. Can you reconstruct it from the spectrum of the wave
operator? On the other hand the dimension of spacetime is a local
property. It can be retrieved from the asymptotic behaviour of the
spectrum of the Dirac operator for large eigenvalues. For compact
spacetime
$M$ this spectrum is discrete. Let us order the
eigenvalues, $...\lambda _{n-1}\leq\lambda _n\leq\lambda
_{n+1}...$ Then Weyl's spectral theorem states that the eigenvalues
grow asymptotically as
$n^{1/{\rm dim}M}$. To explore a local property of spacetime we
only need the high energy part of the spectrum. This is in nice
agreement with our intuition from quantum mechanics and
motivates the name spectral triple.

Differential forms are the main local ingredient of the Yang-Mills
Lagrangian. They too are reconstructed from the spectral triple
using the Dirac operator.  For example the 1-form
$\de a$ for a function $a$ on spacetime is reconstructed as
$[\ddd,\rho (a)]$. This is again motivated from quantum
mechanics. Indeed in a 1+0 dimensional spacetime $\de a$ is just
the time derivative of the `observable' $a$ and is associated to the
commutator of the Hamilton operator with $a$. 

Finally and most
importantly for us, Einstein's derivation of general relativity from
Riemannian geometry can be extended to spectral triples. 
Einstein's starting point is Newton's equation which describes the
dynamics of a point particle. Since in noncommutative geometry
points lose their meaning, Connes' starting point is the Dirac
equation which describes the dynamics of a quantum particle.
Thereby Connes' derivations remains valid for all
 spectral triples, commutative or not.  In accordance with our
language from quantum mechanics, a noncommutative spectral
triple is addressed as noncommutative space or noncommutative
geometry. Of course we are eager to see what Einstein's derivation
becomes in a noncommutative geometry. The simplest such
geometry describes a direct product of a four dimensional
spacetime with a discrete space of points. In other words we are
talking about  Kaluza-Klein  models where the transverse space is
of dimension zero. Indeed one of the advantages of the description
of geometry by spectral triples, commutative or not, is that
continuous and discrete spaces are included in the same formalism.
Connes and Chamseddine
\cite{tresch, grav, cc} have repeated Einstein's derivation for
these discrete Kaluza-Klein geometries. The result is absolutely
amazing. Starting from the free Dirac Lagrangian alone they
derive the Einstein-Hilbert Lagrangian of gravity and
simultaneous they get for free  the Yang-Mills
Lagrangian, the Klein-Gordon Lagrangian, the Higgs potential,
the covariant Dirac Lagrangian and the Yukawa terms. In other
words they derive the entire slot machine of Yang-Mills-Higgs
from geometry. In this geometry the Higgs scalar is a 1-form just
as the gauge bosons. The latter define parallel transport in the four
continuous directions of spacetime, the former defines parallel
transport in the discrete direction. The Yukawa terms are
the minimal couplings of the scalars, and the scalar self coupling,
$\lambda $ is related to the gauge boson self coupling $g^2$ in the
nonAbelian case. In these noncommutative spaces of discrete
Kaluza-Klein type the uncertainty is transverse, in the sense that
in the four dimensional Riemannian space, points are still sharp.
Take for example for the transverse dimension the two-point
space. Then the direct product is the `two sheeted universe'
consisting of two identical copies of the four dimensional
Riemannian space. While each point of the Riemannian space is
well localized the uncertainty is that you do not know on which
copy you are. The distance between the two copies is
measured by the Higgs field. The two sheeted universe \cite{dkm}
was one of the first noncommutative examples to exhibit 
spontaneous symmetry breaking.

\begin{figure}[h]
\label{slot3}
\epsfxsize=8cm
\hspace{4.1cm}
\epsfbox{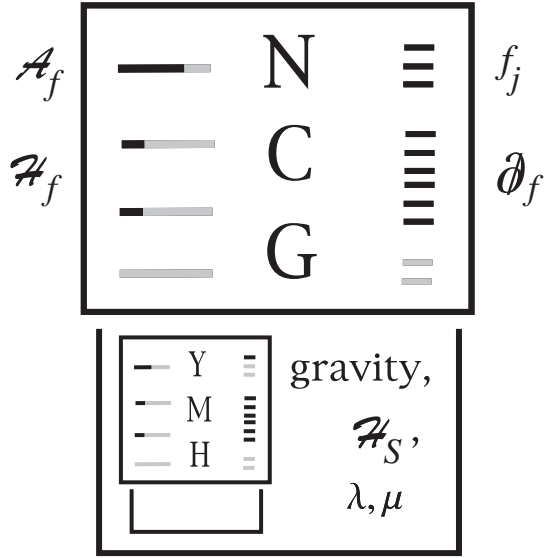}
\caption{Connes' slot machine}
\end{figure}

Coming back to the slot machine, the only arbitrary input left is
the choice of the spectral triple $(\aa_f,\hh_f,\ddd_f)$ describing
the discrete space and three constants, $f_0,f_2,f_4$, figure
3. We use the index
$\cdot_f$ for finite because discrete spaces are zero-dimensional.
In accordance with Weyl's theorem the algebra $\aa_f$ and the
representation space
$\hh_f$ are both finite dimensional. The classification of those is
well known, the algebra is a sum of matrix algebras with the
fundamental or singlet representations. The compact group $G$
from the input is the group of unitaries,
$U(\aa_f)$ and the fermionic representations are $\hh_f
=\hh_L\op\hh_R$. The discrete Dirac operator $\ddd_f$ is the
fermionic mass matrix, fermion masses and mixings. The
Yukawa couplings remain input but they are constrained by the
axioms of the spectral triple. These constraints are so tight that
only very few Yang-Mills-Higgs models can be derived from
noncommutative geometry as pseudo forces. No left-right
symmetric model can \cite{florian}, no Grand Unified Theory can,
for instance the $SU(5)$ model needs a 10-dimensional fermion
representations, $SO(10)$  16-dimensional ones, $E_6$ is not the
group of an associative algebra. Moreover the last two models are
left-right symmetric. Much effort has gone into the construction
of a supersymmetric model from noncommutative geometry, in
vain \cite{kw}.

On the output side we 
find of course gravity, its cosmological constant is related to the
input parameter
$f_0$, Newtons constant to $f_2$. We find
 the complete Yang-Mills-Higgs Lagrangian
(\ref{ymh}). Its Higgs sector, the representation $\hh_S$ and the
potential
$V(\varphi )$, is entirely computed from the data of the finite
spectral triple.  The
Higgs self coupling $\lambda $ is related to the gauge
coupling $g$. Both are computed from $f_4$.

\begin{figure}[h]
\label{versus}
\epsfxsize=11cm
\hspace{2.1cm}
\epsfbox{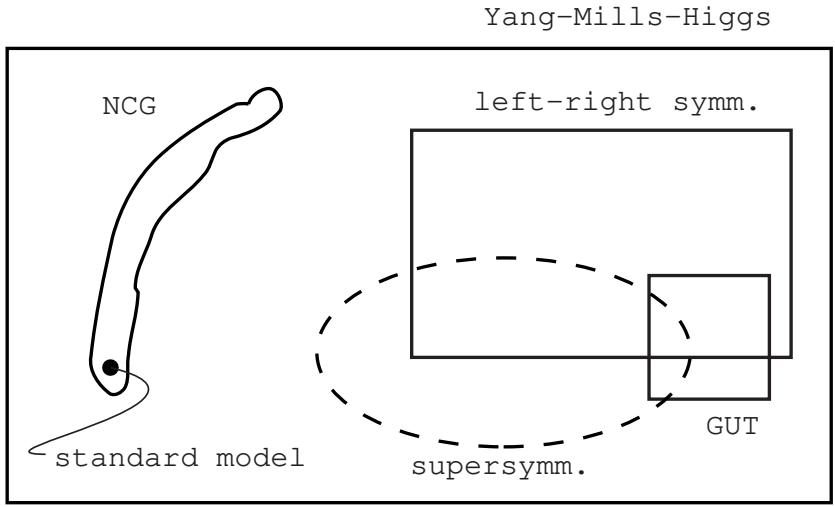}
\caption{Pseudo forces from noncommutative geometry}
\end{figure}

The standard model fits perfectly into this frame, see
figure 4. Indeed you check that its group consists of
unitaries, equation (\ref{group}), and that its fermionic
representation consists of fundamental and singlet
representations, equations (\ref{hl}) and (\ref{hr}). Furthermore
the computation of the scalar representation
$\hh_S$ yields equation (\ref{hs}) on the nose. Not enough, the
six intriguing properties of the standard model listed in
subsection \ref{win} are ad hoc choices in the frame of the
Yang-Mills-Higgs slot machine, they derive from the axioms of the
spectral triple together with the physical assumption that parity is
violated. In particular, the fermionic hypercharges can be
computed and come out correctly \cite{fare}.  Finally the relations
among coupling constants read in the standard model,
\bb g_2^2=g_3^2=3\lambda .\ee
If, like in Grand Unified Theories, we add the hypothesis of the big
desert then standard renormalization flow gives a unification scale
of $\Lambda =10^{17} $ GeV where the uncertainty of spacetime is
expected to become longitudinal and consequently the coupling
constants should cease to run. At the same time we get a
Higgs mass of
$m_H=171\,\pm\,5$ GeV for a top mass of $174.3\,\pm\,5.1$
GeV.

In \cite{lit} you may find additional references on
noncommutative geometry and its applications to forces. I
recommend particularly the recent Costa Rica book \cite{cr}.

\section{Outlook}

Amelino-Camelia gives three
 arguments \cite{ac} that the experimental
observation of the uncertainty at $10^{17}$ Gev or $10^{-40}$ s 
might be possible within the next ten years. These observations
concern local experiments on earth, like gravitational wave
detectors, and measurements at cosmological scale, like $\gamma $
ray bursts.

We are optimistic to be able to single out the standard model within
its noncommutative frame by  one physical requirement: that the
spontaneous symmetry breaking on which we have no 
handle be such that it allow different fermion masses within one
irreducible multiplet, like the left-handed top and bottom quarks,
that sit in the same isospin doublet.

Noncommutative geometry reconciles Riemannian geometry and
uncertainty. We expect the new paradigm, that does not recognize
short distances, to clean up  quantum field theory and to reconcile
it with general relativity. Progress in this direction exists: Connes,
Moscovici and  Kreimer discovered a subtle
link between a noncommutative generalization of the index
theorem and perturbative quantum field theory. This link is a
Hopf algebra relevant to both theories \cite{cmk}.

 \end{document}